\documentclass[a4paper,11pt]{article}
\usepackage{pos}
\usepackage{listings}
\usepackage{natbib}
\setcitestyle{square,numbers,sort&compress}

\definecolor{codegreen}{rgb}{0,0.6,0}
\definecolor{codegray}{rgb}{0.5,0.5,0.5}
\definecolor{codepurple}{rgb}{0.58,0,0.82}
\definecolor{backcolour}{rgb}{0.95,0.95,0.92}

\lstdefinestyle{mystyle}{
    backgroundcolor=\color{backcolour},   
    commentstyle=\color{codegray},
    keywordstyle=\color{codegreen},
    numberstyle=\tiny\color{codegray},
    stringstyle=\color{orange},
    basicstyle=\ttfamily\footnotesize,
    breakatwhitespace=false,         
    breaklines=true,                 
    captionpos=b,                    
    keepspaces=true,                 
    numbers=left,                    
    numbersep=5pt,                  
    showspaces=false,                
    showstringspaces=false,
    showtabs=false,                  
    tabsize=2
}

\title{Prometheus: An Open-Source Neutrino Telescope Simulation}
 \ShortTitle{Prometheus: An Open-Source Neutrino Telescope Simulation}

\author*[a]{David Kim for the Prometheus authors}

\affiliation[a]{
    Department of Physics, Cornell University, Ithaca 14853, NY, United States
    }

\emailAdd{dsk265@cornell.edu}

\abstract{
The soon-to-be-realized, global network of neutrino telescopes will allow new opportunities for collaboration between detectors.
While each detector is distinct, they share the same underlying physical processes and detection principles.
The full simulation chain for these telescopes is typically proprietary which limits the opportunity for joint studies. This means there is no consistent framework for simulating multiple detectors.
To overcome these challenges, we introduce Prometheus, an open-source simulation tool for neutrino telescopes.
Prometheus simulates neutrino injection and final state and photon propagation in both ice and water.
It also supports user-supplied injection and detector specifications.
In this contribution, we will introduce the software; show its runtime performance; and highlight successes in reproducing simulation results from multiple ice- and water-based observatories.
}

\ConferenceLogo{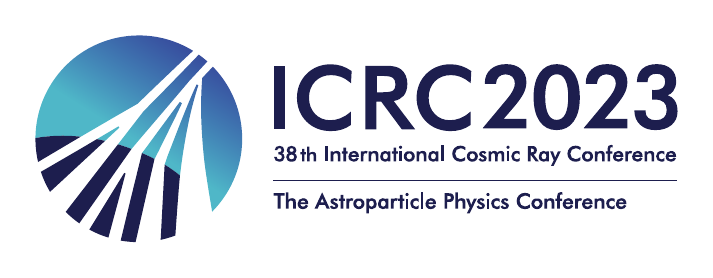}

\FullConference{%
The 38th International Cosmic Ray Conference (ICRC2023)\\
  26 July -- 3 August, 2023\\
  Nagoya, Japan}

\begin{document}
\maketitle

\section{Introduction}
\label{sec:intro}

\begin{figure}[b]
  \centering
  \includegraphics[width=0.96\textwidth]{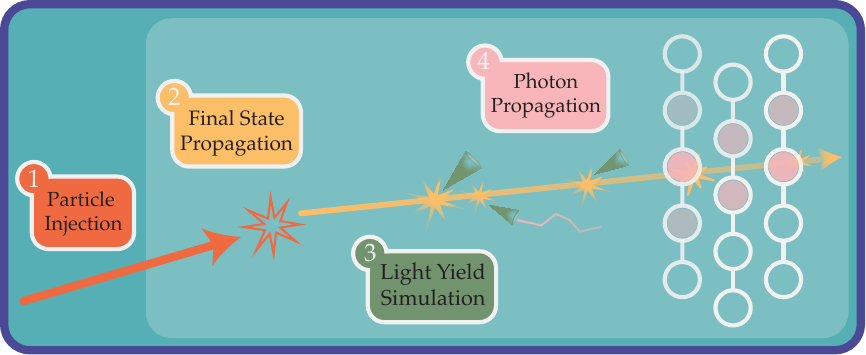}
  \caption{
    \textbf{\textit{Schematic showing the physical processes \texttt{Prometheus} models.}}
    (1), \texttt{Prometheus} selects an interaction vertex within \textit{simulation volume}, depicted here by the lighter-colored region.
    (2), the final states of this interaction are then propagated, accounting for energy losses and any daughter particles which may be produced.
    (3), these losses are then converted to a number of photons.
    (4), finally, these photons are then propagated until they either are absorbed or reach an optical module.
    }
  \label{fig:physics_process}
\end{figure}

The global network of neutrino telescopes, defined as gigaton-scale neutrino detectors, has allowed us observe the Universe in new ways. 
The subset of this network of telescope deployed on ice or water includes the IceCube Neutrino Observatory~\cite{IceCube:2016zyt} near the South Pole, proposed detectors ORCA and ARCA~\cite{KM3Net:2016zxf} in the Mediterranean Sea (KM3NeT collaboration), and Baikal-GVD in Lake Baikal, Russia~\cite{Avrorin:2015wba} (BDUNT collaboration). 
Additionally, new experiments like P-ONE~\cite{P-ONE:2020ljt} off the coast of Vancouver and TRIDENT~\cite{Ye:2022vbk} in the South China Sea are underway, as well as an expansion for the IceCube Observatory~\cite{Ishihara:2019aao, IceCube-Gen2:2020qha}.

These telescopes all share many technological features.
Each of these detectors operate by detecting Cherenkov photons emitted by neutrino interaction byproducts, and as such follow the same general simulation chain illustrated in Fig. 1. 
The only proprietary step is the final detector response, which occurs after an optical module (OM) has detected a photon. 
Yet for many years we have lacked a simulation framework that takes advantage of this similarity.
Existing packages individually cover one or two of these common steps, but until now there has been no easy way to simulate a particle from injection to photon propagation. 

\texttt{Prometheus}~\cite{Lazar:2023prm} looks to correct this by providing an integrated framework to simulate these common steps for arbitrary detectors in ice and water, using a combination of publicly available packages and those newly developed for this work.
Neutrino injection is handled by \texttt{LeptonInjector}\cite{IceCube:2020tcq}, an event generation recently developed by the IceCube Collaboration.
Taus and muons are then propagated by \texttt{PROPOSAL}~\cite{Koehne:2013gpa}. 
Light yield simulation and photon propagation in ice relies on \texttt{PPC}~\cite{chirkin2022kpl}, while in water these steps are covered by \texttt{Fennel}~\cite{fennel2022@github} and \texttt{Hyperion}, respectively.

\texttt{Prometheus}'s flexibility allows one to optimize detector configurations for specific physics goals, while the common format allows one to develop reconstruction techniques that may be applied across different experiments.
With the recent explosion in machine-learning research, it is now more important than ever that we are able to rapidly implement and test new ideas without relying on tools and data that may be proprietary to their experiments. 

The rest of this article is organized as follows. 
In Sec.~\ref{sec:validation} we outline the format of \texttt{Prometheus}'s output and validate against published results. 
In Sec.~\ref{sec:community} we present community work that employs \texttt{Prometheus}. 
In Sec.~\ref{sec:examples} we provide a short example running \texttt{Prometheus} code. 
Finally, in Sec.~\ref{sec:conclusion} we conclude and offer our future outlook.
\section{\texttt{Prometheus}: Output and Validation}
\label{sec:validation}

\begin{figure*}[t]
  \centering
  \includegraphics[width=1\textwidth]{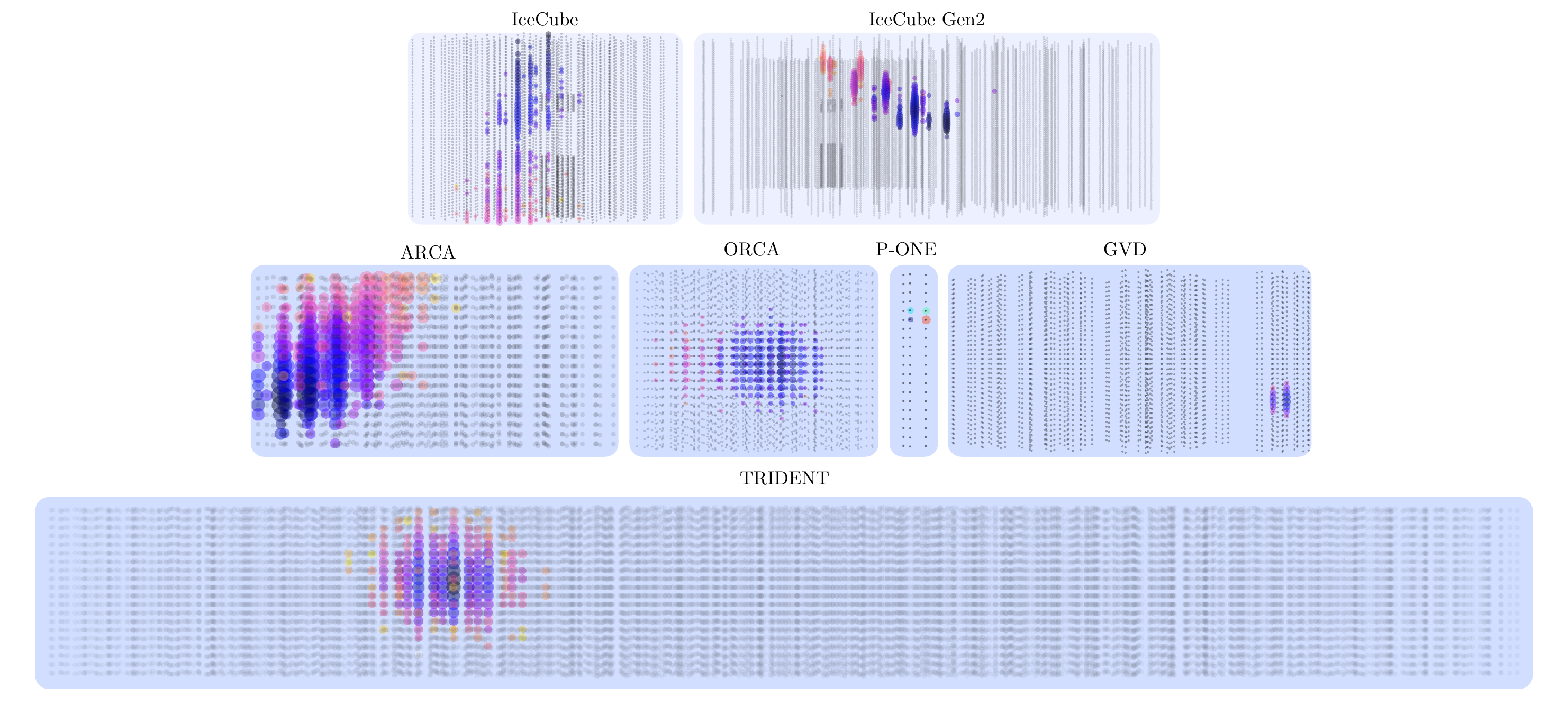}
  \caption{\textbf{\textit{Event views for various detector geometries.}}
  This shows the events created by either $\nu_{\mu}$ charged-current or $\nu_{e}$ charged-current interactions in a variety of geometries of current and proposed neutrino telescopes.
  Each black dot is an OM, while each colored dot indicates the average time at which photons arrived at the OM; black indicates an earlier arrival, orange indicates a later arrival, and purple an arrival in between.
  Furthermore, the size of the colored spheres is proportional to the number of photons that arrived at the OM.
  Detectors which appear against lighter blue backgrounds---the top row---are ice-based, while those against the darker blue backgrounds are water-based.
  }
\end{figure*}

\texttt{Prometheus} outputs to \texttt{Parquet}~\cite{parquet_docs} files that include two primary fields—\texttt{photons} and \texttt{mc\_truth}.
\texttt{photons} contains information on photons that produce hits in user-defined detection regions.
This includes photon arrival time, OM identification numbers, OM position, and the final-state particle that produced the photon. 
Fig. 1 shows event displays for various detectors generated using the information in \texttt{photons}.
\texttt{mc\_truth} includes information on the injection like the interaction vertex; the initial neutrino type, energy, and direction; and the final state types, energies, directions, and parent particles.
Users may also save the configuration file (see Sec.~\ref{sec:examples}) as a \texttt{json} file. 
This allows the user to resimulate events using the same parameters, which is useful for comparing the same event across multiple detector geometries.

\begin{figure}[t]
  \centering
  \includegraphics[width=\textwidth]{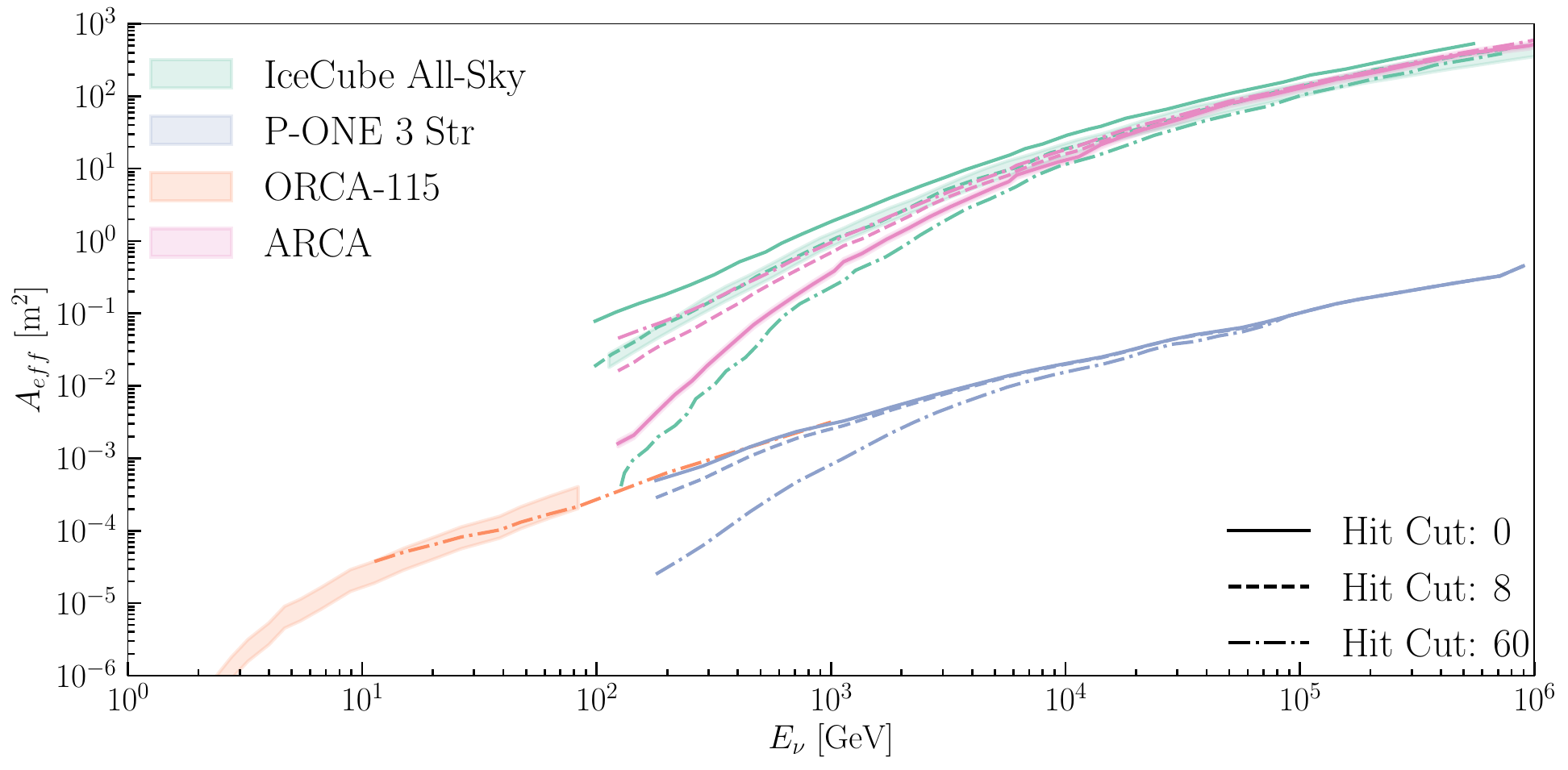}
  \caption{
  \textbf{\textit{Effective area computed using \texttt{Prometheus} with comparisons to published results.}}
  We compare the $\nu_{\mu}$ effective areas computed with \texttt{Prometheus} for IceCube, P-ONE3, ORCA, and ARCA for three different hit requirements, denoted by different line styles, to published effective areas.
  The IceCube effective area, taken from Ref.~\cite{karle:2009abc}, is for $\nu_{\mu}+\nu_{\tau}$ events which pass the SMT-8 trigger and agrees with our calculation to within uncertainties.
  The ARCA~\cite{KM3Net:2016zxf} and ORCA~\cite{deWasseige:2020dnq} cases effective areas are constructed with more complicated hit requirements.
  Still, the scale and shape of the ORCA and ARCA effective areas and the \texttt{Prometheus} effective areas agree within uncertainties despite the simplified selection criterion.
  As of the publication of this proceeding, there is no published effective area for P-ONE3.
  }
  \label{fig:effa}
\end{figure}

The information stored in \texttt{mc\_truth} allows us to compute effective areas when combined with the weights from \texttt{LeptonWeighter}. 
Since it mainly depends on the physics implemented in \texttt{Prometheus}---such as neutrino-nucleon interactions, lepton range, and photon propagation---effective area serves as a reliable indicator of our code's performance when all simulation steps are properly integrated.
We can then validate our simulation by comparing these effective areas to published data.
Fig.~\ref{fig:effa} compares different experiments published effective areas to our estimation using \texttt{Prometheus} simulations. 
It is worth noting that the calculations for effective area rely on detector-specific cuts and OM response, to which we have limited or no access. 
We can therefore expect differences of $\mathcal{O}(10\%)$ from these missing detector details.

\section{Community Contributions}
\label{sec:community}

\begin{figure}[t]
\centering
\includegraphics[width=0.45\textwidth]{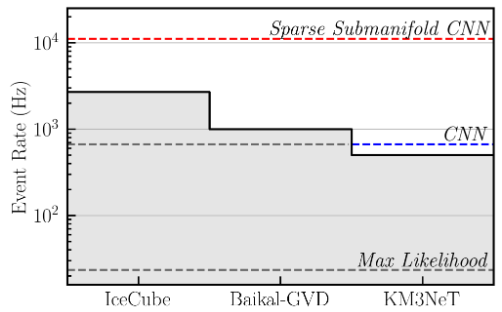}
\caption{
  \textbf{\textit{\cite{IceCube:2016xci, Bakker:2011, AVRORIN201130} Event rates of triggers in different neutrino telescopes compared to the run-times of various reconstruction methods.}} 
  Notably, sparse submanifold CNNs can process events well above standard trigger rates in both ice- and water-based experiments. The CNN and
  maximum likelihood method run-times are taken from~\cite{Hunnefeld:2017}.
  Reproduced with permission of the authors from Ref.~\cite{Yu:2023pos}.
  }
\label{fig:event_rates}
\end{figure}

\begin{figure}[t]
\centering
\includegraphics[width=0.5\textwidth]{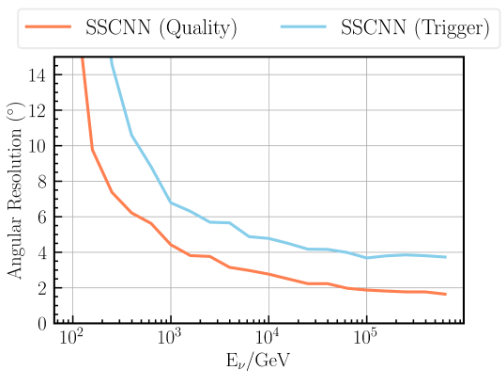}
\caption{
  \textbf{\textit{Angular reconstruction performance as a function of the true neutrino energy.}}
  The angular resolution results are binned by the true neutrino energy, with the median taken from each bin to form the lines shown~\cite{Yu:2023pos}.
  }
\end{figure}
\label{fig:angular_resolution}

Developing a published reconstruction is often a long-term effort internal to an experiment, meaning methods are slow to implement and difficult to compare.
\texttt{Prometheus} remedies this by allowing users to readily generate the large data sets necessary to test the feasibility and relative performance of different reconstruction methods.
In this section we highlight two submissions at ICRC 2023 that utilize \texttt{Prometheus} in such a way.
The first of these works is a software-focused effort that looks to improve the speed of first reconstructions at detectors and the second is a hardware-focused effort concerning GPU alternatives for low power computing. 

Ref.~\cite{Yu:2023pos}, proposes sparse-submanifold convolutions (SSCNNs) as an alternative to the convolutional neural network (CNN) and traditional trigger-level event reconstructions currently used in neutrino telescopes. 
Their directional and energy algorithm is trained on a data set of 412892 events and tested on a further 50000 events, cut from a set of 3 million generated by \texttt{Prometheus}. 
As seen in Fig.~\ref{fig:event_rates}, the SSCNN is capable of running at speeds comparable to the neutrino telescope trigger rates.
In Fig. 5, it achieves a median angular resolution below $4^\circ$ for the highest-energy trigger-level events, which matches or outperforms currently employed reconstructions.
Their work could be applied to improve on-site reconstructions and filtering for notable events at any detector. 

Ref.~\cite{Jin:2023}, is another \texttt{Prometheus} user submission that explores hardware accelerators to improving event reconstruction. 
Specifically, they look at how a Tensor Processing Unit (TPU) compatible algorithm could lower energy consumption while running comparable operations to the current GPU-based ones. 
Their model reaches a median angular resolution around $5^\circ$ above $10^3$ GeV. 
Meanwhile, approximate peak total power consumption drops from 100W on the best performing GPU (Apple M1 Pro chip) down to 3W on the Google Edge TPU.
The portion of the power consumption directed to the ML accelerator component drops from 15W to 2W.
This submission illustrates \texttt{Prometheus}'s utility in testing more speculative ideas and providing proof-of-concepts to encourage further work, particularly in the direction of machine learning.

The findings of both works are generic to any ice or water neutrino telescope.
Their models are trained and tested on example detector geometries rather than the geometry of any existing or proposed detector.
Since \texttt{Prometheus} is fully open-source, we hope it will facilitate not only easier iteration on these techniques but also greater collaboration in sharing and adapting them.

\section{Examples}
\label{sec:examples}

In this section we will walk through producing a simulation of $\nu_{\mu}$ charged-current interactions in an ice-based detector.
This example will use mostly default values for injection parameters to show the essential steps in running a simulation, after which we will show how the user can set these parameters. 

In this example, we set the number of events to simulate, the detector geometry file, and---using the final state particles---the event type.
With just this, we can simulate events with \texttt{Prometheus}!

\begin{lstlisting}[language=python,upquote=true]
import prometheus
from prometheus import config, Prometheus

config["run"]["nevents"] = 100
geofile = f"{resource_dir}/geofiles/demo_ice.geo"
config["detector"]["geo file"] = geofile
injection_config["simulation"]["final state 1"] = "MuMinus"
injection_config["simulation"]["final state 2"] = "Hadrons"

p = Prometheus(config)
p.sim()
\end{lstlisting}

As briefly shown here, the \texttt{config} dictionary is our primary interface for configuring \texttt{Prometheus}. 
Key parameters not shown here include the output directory; random state seed; and injection parameters like injection angle and energy.
Information on a detector's medium, either ice or water, is stored in its geometry file.
The \texttt{geofiles} directory in the GitHub repository has geometry files for all of the detectors in Fig. 2. 
Again, this only scratches the surface of \texttt{Prometheus}'s capabilities.
For a more thorough description of all the options and features available see Ref.~\cite{Lazar:2023prm}, which has more in-depth examples for $\nu_{\mu}$ charged-current events in ice and $\bar\nu_{e}$ neutral-current events in water as well as directions on constructing a detector, weighting events, and getting event rates.

\section{Conclusion}
\label{sec:conclusion}

In this submission we have introduced \texttt{Prometheus} as an open-source software package for simulating  neutrino telescopes. 
We have provided a brief example for simulating events in an ice-based detector and highlighted two current works that employ \texttt{Prometheus}. 
\texttt{Prometheus}'s flexibility of input for detector geometry and injection parameters allows it to handle simulation for the full range of existing and proposed telescopes in both water and ice.

As we have demonstrated in the highlighted community contributions, \texttt{Prometheus} facilitates the implementation of new ideas without the need for proprietary data, or for data on a scale not yet available.
Via its particular application to machine-learning models, \texttt{Prometheus} can be a key piece in accelerating the development of faster, more efficient reconstructions for all detectors.

Finally, it is our hope that \texttt{Prometheus} opens the door for greater collaboration within the community. 
By encouraging the sharing of methods and simulated data sets, we hope work done by any one effort more quickly and easily becomes progress for every group in the global neutrino telescope network.

\section{Acknowledgements}

We would like to thank all users who tested early versions of this software, including---in no particular order---Miaochen Jin, Eliot Genton, Tong Zhu, Rasmus {\O}rs{\o}e, Savanna Coffel, and Felix Yu.
The authors that developed \texttt{Prometheus} were supported by Faculty of Arts and Sciences of Harvard University, the Alfred P. Sloan Foundation, NSF under grants PLR-1600823, PHY-1607644, Wisconsin Research Council with funds granted by the Wisconsin Alumni Research Foundation, Australian Government through the Australian Research Council's Discovery Projects funding scheme (project DP220101727), and Lynne Sacks and Paul Kim.

\bibliographystyle{ICRC}
\bibliography{main}



%
%
%

\end{document}